\documentclass[conference]{IEEEtran}
\IEEEoverridecommandlockouts

\usepackage{cite}
\usepackage{amsmath,amssymb,amsfonts}
\usepackage{algorithmic}
\usepackage{graphicx}
\usepackage{textcomp}
\usepackage{xcolor}
\usepackage{booktabs} 
\usepackage{caption}
\usepackage{lipsum} 
\usepackage{multirow}
\usepackage{tabularx}
\usepackage{hyperref} 

\def\BibTeX{{\rm B\kern-.05em{\sc i\kern-.025em b}\kern-.08em
    T\kern-.1667em\lower.7ex\hbox{E}\kern-.125emX}}

\begin{document}

\title{Quantum Adaptive Excitation Network with\\ Variational Quantum Circuits for\\ Channel Attention 

}

\author{
\IEEEauthorblockN{
    Yu-Chao Hsu\IEEEauthorrefmark{1}\IEEEauthorrefmark{2}, 
    Kuan-Cheng Chen\IEEEauthorrefmark{3}\IEEEauthorrefmark{4},
    Tai-Yue Li \IEEEauthorrefmark{2}
    and Nan-Yow Chen\IEEEauthorrefmark{2}
}
\IEEEauthorblockA{\IEEEauthorrefmark{1} Cross College Elite Program, National Cheng Kung University, Tainan, Taiwan}
\IEEEauthorblockA{\IEEEauthorrefmark{2} National Center for High-Performance Computing, Hsinchu, Taiwan}
\IEEEauthorblockA{\IEEEauthorrefmark{3} Department of Electrical and Electronic Engineering, Imperial College London, London, UK}
\IEEEauthorblockA{\IEEEauthorrefmark{4} Centre for Quantum Engineering, Science and Technology (QuEST), Imperial College London, London, UK}
}

\maketitle

\begin{abstract}
In this work, we introduce the Quantum Adaptive Excitation Network (QAE-Net), a hybrid quantum-classical framework designed to enhance channel attention mechanisms in Convolutional Neural Networks (CNNs). QAE-Net replaces the classical excitation block of Squeeze-and-Excitation modules with a shallow Variational Quantum Circuit (VQC), leveraging quantum superposition and entanglement to capture higher-order inter-channel dependencies that are challenging to model with purely classical approaches. We evaluate QAE-Net on benchmark image classification tasks, including MNIST, FashionMNIST, and CIFAR-10, and observe consistent performance improvements across all datasets, with particularly notable gains on tasks involving three-channel inputs. Furthermore, experimental results demonstrate that increasing the number of variational layers in the quantum circuit leads to progressively higher classification accuracy, underscoring the expressivity benefits of deeper quantum models. These findings highlight the potential of integrating VQCs into CNN architectures to improve representational capacity while maintaining compatibility with near-term quantum devices. The proposed approach is tailored for the Noisy Intermediate-Scale Quantum (NISQ) era, offering a scalable and feasible pathway for deploying quantum-enhanced attention mechanisms in practical deep learning workflows.

\end{abstract}



\section{Introduction}
Convolutional Neural Networks (CNNs)\cite{lecun2002gradient,li2021survey} have revolutionized computer vision by hierarchically extracting features across spatial scales, enabling breakthroughs in tasks such as image classification, object detection, and semantic segmentation. These networks typically stack convolutional and pooling layers to capture local patterns, followed by deeper layers that integrate context over larger receptive fields. Despite their success, standard CNN architectures treat all feature channels equivalently, potentially overlooking subtle inter-channel dependencies that encode critical semantic cues. 
To address this limitation, channel-wise attention mechanisms have been introduced, with Squeeze-and-Excitation Networks (SENet)\cite{hu2018squeeze} leading the way. SENet applies a ``squeeze" operation---global average pooling---to condense spatial information into a channel descriptor, then performs an "excitation" through two fully connected layers with non-linear activations to learn channel relationships.
The resulting attention weights adaptively recalibrate channel responses by emphasizing informative features while suppressing less relevant ones, and have been successfully applied to a wide range of tasks~\cite{luo2024genuine,zhang2024se,qiu2025high}.
Numerous extensions (e.g., ECA-Net~\cite{wang2020eca}, GSoP-Net~\cite{gao2019global}) have further optimized parameter efficiency and higher-order statistics, yet remain constrained by classical function approximators and the increasing computational overhead in deeper networks.

Quantum Computing (QC) offers fundamentally new paradigms for data representation and processing. By leveraging quantum superposition, a register of \(n\) qubits can simultaneously encode \(2^n\) computational basis states, while entanglement enables complex correlations with no classical counterpart. Building on these principles, the field of Quantum Machine Learning (QML)~\cite{schuld2023quantum,wossnig2021quantum,Maria2019quantum} explores the integration of quantum computing techniques with classical machine learning frameworks, with the goal of enhancing both computational efficiency (reducing time complexity) and representational capacity (reducing space complexity). A diverse array of QML models has been proposed, including the Quantum Support Vector Machine (QSVM)~\cite{rebentrost2014quantum,havlivcek2019supervised,chen2023quantum,chen2024validating,ma2025robust,tai2022quantum}, Quantum Kernel-Based Long Short-Term Memory (QK-LSTM) networks~\cite{hsu2025quantum,hsuKernel2025quantum} and Quantum Convolutional Neural Networks (QCNNs)~\cite{oh2020tutorial,cong2019quantum,hur2022quantum,an2025quantum}.

Among these approaches, Variational Quantum Circuits (VQCs)\cite{cerezo2021variational} have emerged as a leading architecture for near-term quantum devices. These circuits consist of trainable single-qubit rotations and entangling gates arranged into shallow ansatzes that can be optimized using classical gradient-based methods. Prior studies have demonstrated the potential of VQCs in tasks such as classification\cite{chen2020hybrid,chen2024novel,chen2025consensus}, generative modeling\cite{leyton2021robust,benedetti2019generative,chen2025differentiable}, quantum-enhanced optimization\cite{moll2018quantum,chen2025learning} and model compression\cite{liu2024quantum,liu2024qtrl,lin2024quantum,liu2025quantum,chen2025quantum}. However, the integration of quantum circuits into internal architectural components—particularly attention mechanisms—remains largely unexplored, presenting new opportunities for hybrid model design.

In this work, we propose the Quantum Adaptive Excitation Network (QAE-Net), which integrates a VQC into the channel attention pipeline of CNNs to form a hybrid quantum-classical module. Instead of classical fully connected layers, QAE-Net encodes the global channel descriptor to a quantum state via angle encoding, processes it through trainable rotations and entangling gates, and measures Pauli-$Z$ observables to produce the quantum feature vector. A lightweight classical linear mapping then expands this vector to channel attention scores, which adaptively recalibrate the original feature maps.

\section{Related work}
Squeeze-and-Excitation Networks (SENet)~\cite{hu2018squeeze} introduced a pioneering channel attention mechanism that enhances representational capacity by explicitly modeling inter-channel dependencies. The SE block consists of a \textit{squeeze} module, which applies global average pooling (GAP) to capture global spatial information, and an \textit{excitation} module, which employs two fully connected (FC) layers with ReLU and Sigmoid activations to produce channel-wise attention weights. 

Subsequent works have extended this framework in various directions. For instance, GSoP-Net~\cite{gao2019global} focuses on improving the squeeze module by incorporating higher-order statistics. Other approaches, such as GCNet~\cite{cao2019gcnet} and SRM~\cite{lee2019srm}, simultaneously enhance both the squeeze and excitation modules to better capture global context and channel interactions.

In contrast to previous approaches, QCA-Net~\cite{zhang2023qca} enhances global spatial information through quantum-inspired mechanisms based on wave function representations. However, QCA-Net cannot be readily integrated into CNNs due to its reliance on hand-crafted features and lack of end-to-end trainability. It also does not employ real quantum circuits. In contrast, our method uses a trainable VQC for excitation, enabling learning and deployment on near-term quantum hardware while effectively modeling channel interactions.

\section{Quantum Preliminaries}\label{sec:qnn}

\begin{figure}[b]
    \centering

    \includegraphics[width=0.5\textwidth]{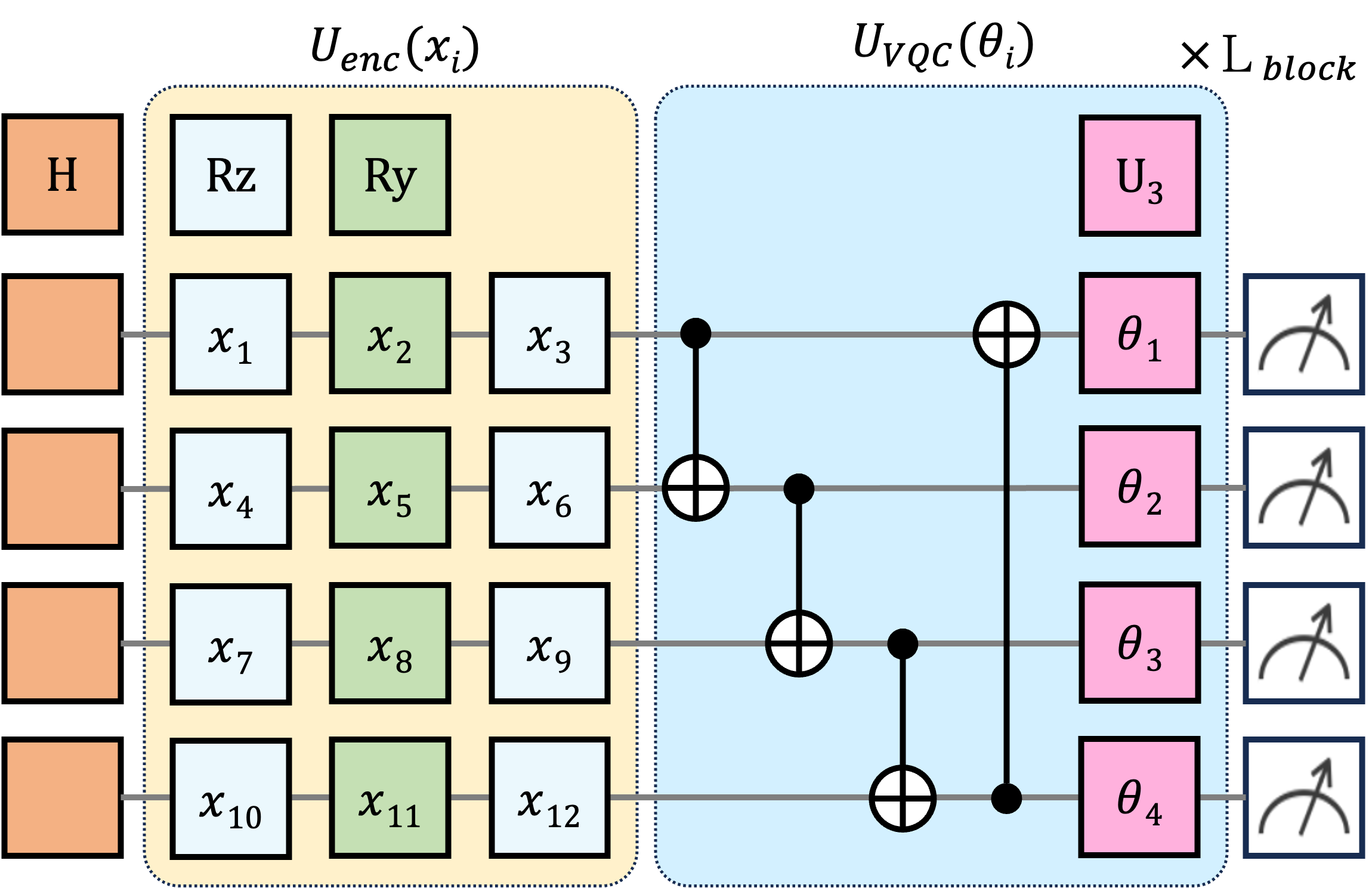}
    \caption{
        Schematic representation of the Quantum Neural Network.
        }
    \label{VQC}

\end{figure}

Quantum Neural Networks (QNNs) are a class of models built upon variational quantum circuits (VQCs), also referred to as parameterized quantum circuits (PQCs) depicted in the Fig.~\ref{VQC}. These circuits consist of trainable quantum gates and entangling operations and are typically implemented on near-term quantum devices. 
The standard design of a QNN typically comprises three main components: (1) an input encoding circuit, (2) parameterized unitary evolution, and (3) quantum measurement. A typical QNN begins by initializing an $n$-qubit quantum register in the ground state ${|0\rangle}^{\otimes n}$. To create an unbiased superposition over all computational basis states, a layer of Hadamard gates is applied to each qubit. 

Classical input data $\boldsymbol{x}$ is then embedded into the quantum system through an encoding circuit. In our work, we adopt angle encoding, where in each qubit is assigned three real-valued inputs to parameterize a full single-qubit SU(2) rotation. The encoding unitary is defined as:
\begin{equation}
    U_{\text{enc}}(\boldsymbol{x}) = \bigotimes_{i=1}^{n} RZ(x_{3i-2}) \cdot RY(x_{3i-1}) \cdot RZ(x_{3i}).
\end{equation}
This encoding operation transforms the initialized superposition state into a data-dependent quantum state, enabling the circuit to capture classical feature information in a quantum representation. Following the encoding circuit $U_{\text{enc}}(\boldsymbol{x})$, the quantum state evolves through $L$ layers of parameterized unitaries:
\begin{equation}
U_{\text{VQC}}(\boldsymbol{\theta}) = \prod_{\ell=1}^{L} \left( U_{\text{ent}} \cdot \bigotimes_{i=1}^{n} U^{(\ell)}_{i}(\theta^{(\ell)}_i) \right),
\end{equation}
where $U^{(\ell)}_{i}(\theta^{(\ell)}_i)$ denotes the single-qubit rotation such as $R_x$, $R_y$, $R_z$, applied to qubit $i$ in the $\ell$-th layer, and $U_{\text{ent}}$ represents the fixed entangling operation applied after each layer of local rotations. In our implementation, we adopt a ring entanglement pattern defined as \(U_{\text{ent}} = \prod_{k=1}^{n-1} \mathrm{CNOT}(k, k+1)\).
The full quantum state is given by:
\begin{equation}
|\psi(\boldsymbol{x}; \boldsymbol{\theta})\rangle = U_{\text{VQC}}(\boldsymbol{\theta}) \cdot U_{\text{enc}}(\boldsymbol{x}) \cdot H^{\otimes n} |0\rangle^{\otimes n}
\end{equation}

\section{Method}\label{sec:qae}

Inspired by the SENet~\cite{hu2018squeeze} and QCA-Net\cite{zhang2023qca}, we propose Quantum Adaptive Excitation Network (QAE-Net) that replaces the classical excitation block with a variational quantum circuit (VQC). The motivation is to capture channel-wise dependencies using the expressive capacity of quantum circuits, while maintaining compatibility with standard convolutional network architectures.
\subsection{Global Information Embedding}
Given an input feature map $\mathbf{X} \in \mathbb{R}^{C \times H \times W}$, we first apply average pooling operation across the spatial dimensions to obtain a channel descriptor $\mathbf{z} \in \mathbb{R}^C$:
\begin{equation}
\mathbf{z}_c = \frac{1}{H \times W} \sum_{i=1}^{H} \sum_{j=1}^{W} \mathbf{X}_{c, i, j}.
\end{equation}
and the channel descriptor is then passed to the VQC for further processing.
\begin{figure*}[!t]
    \centering
    \includegraphics[width=0.95\textwidth]{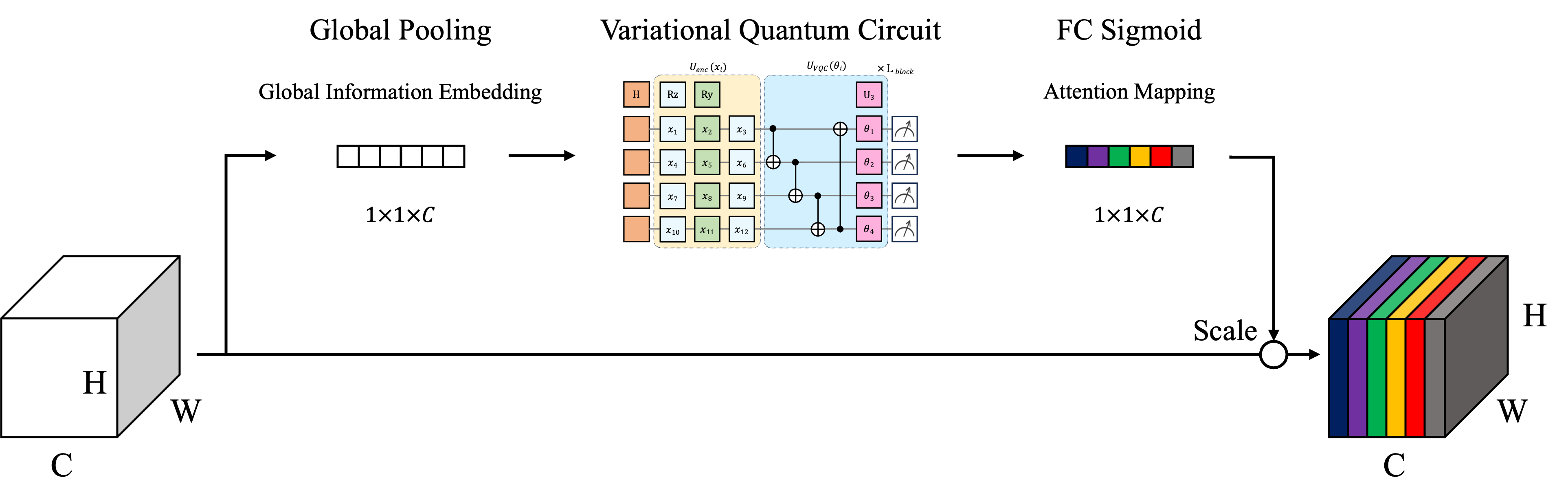}
    \caption{
    Illustration of the Quantum Adaptive Excitation Network.
    }
    \label{QSEnet}
    \vspace{-12pt}
\end{figure*}

\subsection{Variational Quantum Circuit}
Instead of passing $\mathbf{z}$ through two classical fully connected (FC) layers as in SENet, we encode it into a quantum state and process it using a trainable quantum circuit. 
Specifically, the vector $\mathbf{z} \in \mathbb{R}^{C}$ is reshaped into $n$ groups of three real-valued components, where $n$ denotes the number of qubits. Each group is used to parameterize a full single-qubit rotation $U_{\text{enc}}(\mathbf{z})$.

Before encoding, a layer of Hadamard gate is applied to each qubit to prepare an unbiased superposition state. The initialized state then evolves as follows:
\begin{equation}
|\psi(\boldsymbol{z}; \boldsymbol{\theta})\rangle = U_{\text{VQC}}(\boldsymbol{\theta}) \cdot U_{\text{enc}}(\boldsymbol{z}) \cdot H^{\otimes n} |0\rangle^{\otimes n}
\end{equation}
where $U_{\text{VQC}}(\boldsymbol{\theta})$ denotes the $L$-layer parameterized quantum circuit comprising single-qubit rotations and entangling gates as defined in Section~\ref{sec:qnn}.

The final quantum embedding is obtained by measuring the expectation values of Pauli-$Z$ observables:
\begin{equation}
Q_i = \langle \psi(\mathbf{z}; \boldsymbol{\theta}) | Z_i | \psi(\mathbf{z}; \boldsymbol{\theta}) \rangle, \quad i = 1, \dots, n.
\end{equation}
This produces a quantum-derived feature vector $\mathbf{Q} \in \mathbb{R}^{n}$ that replaces the classical intermediate representation used in SENet.
\subsection{Cross-Channel Interaction}
To restore the original channel dimensionality $C$, the quantum output $\mathbf{Q}$ is fed into a classical linear layer to combine all channels linearly without changing the number of channels, followed by a sigmoid activation:
\begin{equation}
\mathbf{S_c} = \sigma(W \cdot\mathbf{Q} + b), \quad \mathbf{S_c} \in \mathbb{R}^C,
\end{equation}
where $\mathbf{Q} \in \mathbb{R}^n$ denotes the quantum-derived feature vector obtained from the $n$-qubit measurement outcomes, $W \in \mathbb{R}^{C \times n}$ is a trainable weight matrix, and $\sigma$ denotes the sigmoid function. This channel-wise attention vector is then applied multiplicatively to the original input:
\begin{equation}
\mathbf{X}'_c = \mathbf{S}_c \cdot \mathbf{X}_c.
\end{equation}

\begin{figure*}[t]
    \centering
    \includegraphics[width=1\textwidth]{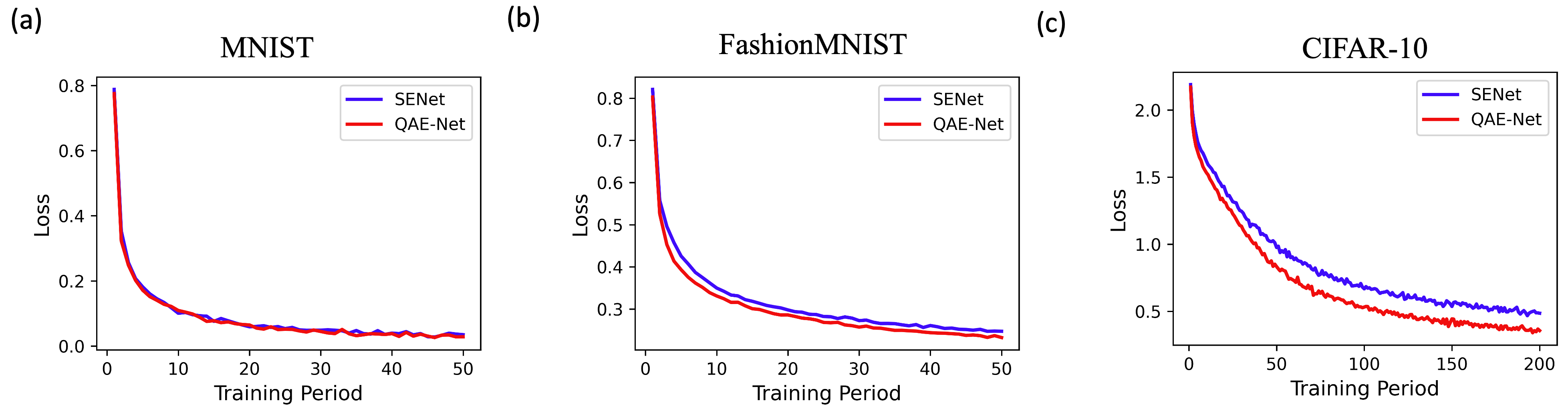}
    \caption{
    Training progression of QAE-Net as visualized through the Cross-Entropy (CE) Loss across epochs for three different
datasets: (a) MNIST, (b) FashionMNIST, and (c) CIFAR-10. } 
    \label{result}
\end{figure*}

\subsection{Optimization Process}

QAE-Net is trained together with the classical convolutional neural network. The entire model, including the variational quantum circuit (VQC), is optimized via gradient-based backpropagation using a cross-entropy loss function for classification.

After the quantum-enhanced channel attention is applied, the recalibrated feature maps $\mathbf{X}'$ are forwarded through the remaining convolutional and fully connected layers for classification. The final layer produces a logit vector $\mathbf{z}^{(i)} \in \mathbb{R}^K$ for each input sample, where $K$ denotes the number of target classes.

The predicted class probabilities are obtained by applying the softmax function:
\begin{equation}
\hat{y}_k^{(i)} = \frac{\exp(z_k^{(i)})}{\sum_{j=1}^{K} \exp(z_j^{(i)})},
\end{equation}
where $z_k^{(i)}$ denotes the unnormalized logit for class $k$ of the $i$-th sample.

The network is trained using the cross-entropy loss, which measures the discrepancy between the predicted distribution $\hat{\mathbf{y}}^{(i)}$ and the one-hot encoded ground truth $\mathbf{y}^{(i)}$:
\begin{equation}
\mathcal{L} = -\frac{1}{N} \sum_{i=1}^{N} \sum_{k=1}^{K} y_k^{(i)} \log \hat{y}_k^{(i)},
\end{equation}
where $N$ is the number of training samples in a mini-batch.
These expectation values constitute the quantum feature vector $\mathbf{Q} \in \mathbb{R}^{n}$, which serves as a replacement for the classical excitation representation in SENet.

In our architecture, the logit vector $\mathbf{z}^{(i)}$ is obtained directly from the measurement outcomes of the quantum circuit, where each component corresponds to an expectation value of a Pauli observable.

To facilitate optimization of the variational quantum circuit, we employ the \textit{parameter-shift rule}~\cite{wierichs2022general}, a standard technique for analytically computing gradients of quantum expectation values with respect to variational parameters. For a quantum observable defined as $Q_i = f(\boldsymbol{z}; \theta_i)$, where $\theta_i$ denotes a tunable gate parameter, the corresponding gradient is given by:
\begin{equation}
\nabla_{\theta_i} Q_i = \frac{1}{2} \left[ f\left(\boldsymbol{z}; \theta_i + \frac{\pi}{2} \right) - f\left(\boldsymbol{z}; \theta_i - \frac{\pi}{2} \right) \right].
\end{equation}
This formulation enables the quantum circuit to be fully differentiable and compatible with classical gradient-based optimizers.

To enhance classification performance, we optimize both the quantum and classical components jointly by minimizing the cross-entropy loss. The hybrid model is trained end-to-end, with gradients propagated through the entire network, including the quantum layers, via the parameter-shift mechanism.

\section{Result}

In this section, we present our approach experiments demonstrating the feasibility of integrating variational quantum circuits into convolutional neural networks via a quantum-enhanced channel attention mechanism. 
We utilized PennyLane\cite{bergholm2018pennylane} and Pytorch\cite{paszke2019pytorch} for our simulation. Quantum circuit optimization was carried out on the \texttt{lightning.gpu} simulator using the parameter-shift rule in conjunction with gradient-based optimization techniques. All experiments were conducted on a system equipped with NVIDIA Tesla V100 GPUs (16GB) and Intel Xeon CPUs.

\subsection{Multi-Class Classification on MNIST, FashionMNIST, and CIFAR-10}
To evaluate the effectiveness of the proposed QAE-Net, we conducted multi-class classification experiments on three widely-used benchmark datasets: MNIST, FashionMNIST, and CIFAR-10, with 60,000 training images and 10,000 testing images.
For each dataset, we compared the performance of a baseline SENet and our quantum-enhanced QAE-Net under identical architectural and training settings.

Both SENet and QAE-Net share the same convolutional backbone for fair comparison. The architecture begins with a convolutional layer (kernel size 5) with 12 channels, followed by a channel attention block (SE or QAE), a second convolutional layer with 16 output channels, and two fully connected layers with hidden dimensions of 256 and 128, respectively. Dropout is applied before classification. In QAE-Net, the classical excitation block in the SE module is replaced with a variational quantum circuit (VQC) consisting of a single variational layer, while all other components of the network architecture remain unchanged.

\begin{table}[!t]
\centering
\caption{Comparative Performance of SENet and QAE-Net Across Benchmark Datasets (Ch.\ = Channels). All QAE-Net configurations employed 4 qubits.}
\label{tab:method_compare_all}
\small
\begin{tabular}{lccccc}
\toprule
\textbf{Dataset} & \textbf{Method} & \textbf{Ch.} & \textbf{Epoch} & \textbf{Params} & \textbf{Acc.\ (\%)} \\
\midrule
\multirow{2}{*}{MNIST} 
    & SENet~\cite{hu2018squeeze} & 1 & 50  & 39,602 & 97.9 \\
    & QAE-Net                    & 1 & 50  & \textbf{39,366}          & \textbf{98.0} \\
\midrule
\multirow{2}{*}{F-MNIST} 
    & SENet~\cite{hu2018squeeze} & 1 & 50  & 39,602 & 91.0 \\
    & QAE-Net                    & 1 & 50  & \textbf{39,366}          & \textbf{91.3} \\
\midrule
\multirow{2}{*}{CIFAR-10} 
    & SENet~\cite{hu2018squeeze} & 3 & 200 & 142,634         & 76.72 \\
    & QAE-Net                    & 3 & 200 & \textbf{142,570} & \textbf{89.08} \\
\bottomrule
\end{tabular}
\end{table}

Experimental results across three benchmark datasets are summarized in Table~\ref{tab:method_compare_all}. QAE-Net consistently outperforms SENet in classification accuracy on MNIST and FashionMNIST while maintaining a comparable number of parameters. Preliminary results on CIFAR-10 also demonstrate the scalability of our quantum-enhanced attention mechanism to more complex image classification tasks under the same network architecture.

Fig.~\ref{result} further illustrates the training dynamics of SE-Net and QAE-Net in terms of cross-entropy (CE) loss over training epochs. As shown in the Fig~\ref{result} , QAE-Net consistently achieves faster convergence and lower final loss values across all three datasets. These results indicate that replacing the classical excitation block with a variational quantum circuit not only improves classification accuracy, but also facilitates more efficient training dynamics.

Together, the quantitative results in Table~\ref{tab:method_compare_all} and the qualitative training progression in Fig.~\ref{result} validate the effectiveness of our proposed QAE-Net. The model demonstrates consistent advantages over its classical counterpart without incurring significant computational overhead, and remains compatible with near-term quantum devices through the use of shallow variational circuits.
\subsection{Effect of Different Number of Variational Layers}
In this section, we use the CIFAR-10 dataset as a benchmark to investigate the impact of different layer
depths of variational quantum circuits on model performance.
Specifically, we examine whether increasing the number of variational  layers enhances the model’s ability to capture complex data representations and improve classification accuracy.

\begin{figure}[!t]
    \centering   \includegraphics[width=0.45\textwidth]{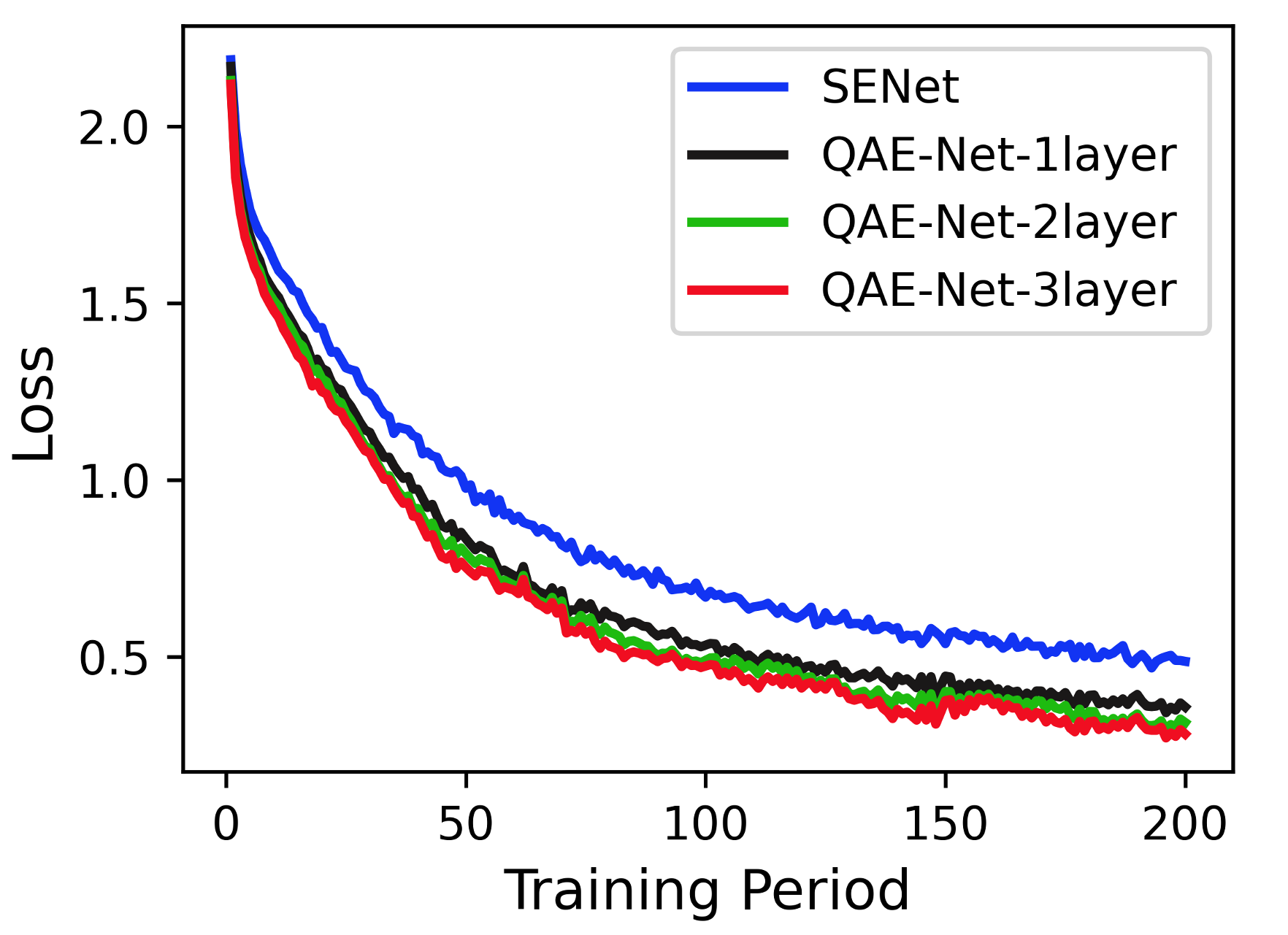}
    \caption{Training progression of QAE-Net with different number of variational layers on CIFAR-10 Dataset}
    \label{result_layer}
\end{figure}

By incrementally increasing
the number of variational layers, we aim to evaluate the trade-off between model expressivity and computational overhead. Fig.~\ref{result_layer} illustrates the training progression of QAE-Net with different numbers of variational layers on the CIFAR-10 dataset. The Fig.~\ref{result_layer} shows the loss values over 200 training epochs for SENet and QAE-Net models with 1, 2, and 3 variational layers. As the number of layers increases, the loss consistently decreases, indicating that deeper quantum models are more effective in learning complex features from the data.

\begin{table}[!b]
\centering
\caption{Performance of QAE-Net with Varying Numbers of Variational Layers}
\label{tab:performance_comparison}
\begin{tabular}{lcccc}
\toprule
\textbf{Metric} 
& \textbf{SENet} 
& \begin{tabular}[c]{@{}c@{}}QAE-Net\\(1 layer)\end{tabular} 
& \begin{tabular}[c]{@{}c@{}}QAE-Net\\(2 layers)\end{tabular} 
& \begin{tabular}[c]{@{}c@{}}QAE-Net\\(3 layers)\end{tabular} \\
\midrule
Epochs & 200 & 200 & 200 & 200 \\
Accuracy (\%) & 76.72 & 89.08 & 90.10 & \textbf{92.30} \\
Qubits & -- & 4 & 4 & 4 \\
Learning Rate & 0.001 & 0.001 & 0.001 & 0.001 \\
\bottomrule
\end{tabular}
\end{table}

As shown in Table  \ref{tab:performance_comparison}, increasing the number of variational layers leads to a consistent improvement in classification accuracy. Specifically, QAE-Net with 1 variational layer achieves an testing accuracy of 89.08\%, which increases to 90.1\% with 2 layers, and further improves to 92.3\% with 3 layers. These results indicate that deeper quantum circuits can better capture complex features in the data, leading to higher classification performance.

\section{Discussion}
In this work, we introduced QAE-Net, a quantum-enhanced channel-attention module that replaces SENet’s excitation block with a shallow Variational Quantum Circuit (VQC), leveraging quantum superposition and entanglement to capture richer inter-channel dependencies. Evaluated on MNIST, Fashion-MNIST, and CIFAR-10, QAE-Net consistently outperforms its classical counterpart, achieving the largest gains on the RGB CIFAR-10 dataset. Increasing the number of variational layers from one to four yields an almost linear improvement in classification accuracy, underscoring the expressivity benefits of deeper quantum circuits.

While variational quantum circuits are known to exhibit barren plateaus—that is, an exponential decay in gradient magnitudes as circuit width and depth increase—the present design mitigates this issue in practice. Specifically, QAE-Net employs a shallow circuit depth and a limited number of qubits, both of which are known to alleviate gradient concentration and improve trainability in near-term settings. Moreover, the hybrid architecture confines the quantum subroutine to auxiliary channel reweighting rather than direct end-to-end classification, reducing the impact of vanishing gradients on the overall optimization process.

Nevertheless, scaling this approach to deeper circuits or larger qubit counts will likely require additional strategies for addressing barren plateaus \cite{mcclean2018barren} and exponential concentration \cite{thanasilp2024exponential} phenomena. Promising directions include problem-inspired ansatz design tailored to the data distribution, layerwise or blockwise training schemes, and the adoption of local cost functions that avoid global optimization over exponentially large Hilbert spaces. As quantum processors mature, QAE-Net provides a scalable pathway for embedding genuine quantum advantages into deep learning architectures, enabling richer representational capacity and potentially more efficient learning in complex computer vision tasks.


\section*{Acknowledgment}
The authors would like to thank the National Center for High-performance Computing of Taiwan for providing computational and storage resources.

\bibliographystyle{ieeetr}
\bibliography{reference} 

\begin{thebibliography}{10}

\bibitem{lecun2002gradient}
Y.~LeCun, L.~Bottou, Y.~Bengio, and P.~Haffner, ``Gradient-based learning applied to document recognition,'' {\em Proceedings of the IEEE}, vol.~86, no.~11, pp.~2278--2324, 2002.

\bibitem{li2021survey}
Z.~Li, F.~Liu, W.~Yang, S.~Peng, and J.~Zhou, ``A survey of convolutional neural networks: analysis, applications, and prospects,'' {\em IEEE transactions on neural networks and learning systems}, vol.~33, no.~12, pp.~6999--7019, 2021.

\bibitem{hu2018squeeze}
J.~Hu, L.~Shen, and G.~Sun, ``Squeeze-and-excitation networks,'' in {\em Proceedings of the IEEE conference on computer vision and pattern recognition}, pp.~7132--7141, 2018.

\bibitem{luo2024genuine}
Y.-J. Luo, X.~Leng, and C.~Zhang, ``Genuine multipartite entanglement verification with convolutional neural networks,'' {\em Physical Review A}, vol.~110, no.~4, p.~042412, 2024.

\bibitem{zhang2024se}
Z.~Zhang, B.~Qin, X.~Gao, T.~Ding, Y.~Zhang, and H.~Wang, ``Se-cnn based emergency control coordination strategy against voltage instability in multi-infeed hybrid ac/dc systems,'' {\em International Journal of Electrical Power \& Energy Systems}, vol.~160, p.~110082, 2024.

\bibitem{qiu2025high}
Y.~Qiu, H.~Huang, Y.~Zhai, Z.~Zheng, and D.~Sun, ``High-fidelity infrared u-shaped residual network digital holography with attention module for crystal growth observation,'' {\em Optics Letters}, vol.~50, no.~6, pp.~1905--1908, 2025.

\bibitem{wang2020eca}
Q.~Wang, B.~Wu, P.~Zhu, P.~Li, W.~Zuo, and Q.~Hu, ``Eca-net: Efficient channel attention for deep convolutional neural networks,'' in {\em Proceedings of the IEEE/CVF conference on computer vision and pattern recognition}, pp.~11534--11542, 2020.

\bibitem{gao2019global}
Z.~Gao, J.~Xie, Q.~Wang, and P.~Li, ``Global second-order pooling convolutional networks,'' in {\em Proceedings of the IEEE/CVF Conference on computer vision and pattern recognition}, pp.~3024--3033, 2019.

\bibitem{schuld2023quantum}
M.~Schuld and F.~Petruccione, ``Quantum machine learning,'' in {\em Encyclopedia of Machine Learning and Data Science}, pp.~1--12, Springer, 2023.

\bibitem{wossnig2021quantum}
L.~Wossnig, ``Quantum machine learning for classical data,'' {\em arXiv preprint arXiv:2105.03684}, 2021.

\bibitem{Maria2019quantum}
M.~Schuld and N.~Killoran, ``Quantum machine learning in feature hilbert spaces,'' {\em Physical review letters}, vol.~122, no.~4, p.~040504, 2019.

\bibitem{rebentrost2014quantum}
P.~Rebentrost, M.~Mohseni, and S.~Lloyd, ``Quantum support vector machine for big data classification,'' {\em Physical review letters}, vol.~113, no.~13, p.~130503, 2014.

\bibitem{havlivcek2019supervised}
V.~Havl{\'\i}{\v{c}}ek, A.~D. C{\'o}rcoles, K.~Temme, A.~W. Harrow, A.~Kandala, J.~M. Chow, and J.~M. Gambetta, ``Supervised learning with quantum-enhanced feature spaces,'' {\em Nature}, vol.~567, no.~7747, pp.~209--212, 2019.

\bibitem{chen2023quantum}
K.-C. Chen, X.~Xu, H.~Makhanov, H.-H. Chung, and C.-Y. Liu, ``Quantum-enhanced support vector machine for large-scale multi-class stellar classification,'' in {\em International Conference on Intelligent Computing}, pp.~155--168, Springer, 2024.

\bibitem{chen2024validating}
K.-C. Chen, T.-Y. Li, Y.-Y. Wang, S.~See, C.-C. Wang, R.~Wille, N.-Y. Chen, A.-C. Yang, and C.-Y. Lin, ``Validating large-scale quantum machine learning: Efficient simulation of quantum support vector machines using tensor networks,'' {\em Machine Learning: Science and Technology}, 2024.

\bibitem{ma2025robust}
W.~Ma, K.-C. Chen, S.~Yu, M.~Liu, and R.~Deng, ``Robust decentralized quantum kernel learning for noisy and adversarial environment,'' {\em arXiv preprint arXiv:2504.13782}, 2025.

\bibitem{tai2022quantum}
T.-Y. Li, V.~R. Mekala, K.-L. Ng, and C.-F. Su, ``Classification of tumor metastasis data by using quantum kernel-based algorithms,'' in {\em 2022 IEEE 22nd International Conference on Bioinformatics and Bioengineering (BIBE)}, pp.~351--354, IEEE, 2022.

\bibitem{hsu2025quantum}
Y.-C. Hsu, T.-Y. Li, and K.-C. Chen, ``Quantum kernel-based long short-term memory,'' in {\em 2025 IEEE International Conference on Acoustics, Speech, and Signal Processing Workshops (ICASSPW)}, pp.~1--5, IEEE, 2025.

\bibitem{hsuKernel2025quantum}
Y.-C. Hsu, N.-Y. Chen, T.-Y. Li, P.-H.~H. Lee, and K.-C. Chen, ``Quantum kernel-based long short-term memory for climate time-series forecasting,'' in {\em 2025 International Conference on Quantum Communications, Networking, and Computing (QCNC)}, pp.~421--426, IEEE, 2025.

\bibitem{oh2020tutorial}
S.~Oh, J.~Choi, and J.~Kim, ``A tutorial on quantum convolutional neural networks (qcnn),'' in {\em 2020 International Conference on Information and Communication Technology Convergence (ICTC)}, pp.~236--239, IEEE, 2020.

\bibitem{cong2019quantum}
I.~Cong, S.~Choi, and M.~D. Lukin, ``Quantum convolutional neural networks,'' {\em Nature Physics}, vol.~15, no.~12, pp.~1273--1278, 2019.

\bibitem{hur2022quantum}
T.~Hur, L.~Kim, and D.~K. Park, ``Quantum convolutional neural network for classical data classification,'' {\em Quantum Machine Intelligence}, vol.~4, no.~1, p.~3, 2022.

\bibitem{an2025quantum}
A.~Ning, T.-Y. Li, and N.-Y. Chen, ``Quantum pointwise convolution: A flexible and scalable approach for neural network enhancement,'' in {\em 2025 International Conference on Quantum Communications, Networking, and Computing (QCNC)}, pp.~371--378, IEEE, 2025.

\bibitem{cerezo2021variational}
M.~Cerezo, A.~Arrasmith, R.~Babbush, S.~C. Benjamin, S.~Endo, K.~Fujii, J.~R. McClean, K.~Mitarai, X.~Yuan, L.~Cincio, {\em et~al.}, ``Variational quantum algorithms,'' {\em Nature Reviews Physics}, vol.~3, no.~9, pp.~625--644, 2021.

\bibitem{chen2020hybrid}
S.~Y.-C. Chen, C.-M. Huang, C.-W. Hsing, and Y.-J. Kao, ``Hybrid quantum-classical classifier based on tensor network and variational quantum circuit,'' {\em arXiv preprint arXiv:2011.14651}, 2020.

\bibitem{chen2024novel}
Y.~Chen, ``A novel image classification framework based on variational quantum algorithms,'' {\em Quantum Information Processing}, vol.~23, no.~10, pp.~1--28, 2024.

\bibitem{chen2025consensus}
K.-C. Chen, W.~Ma, and X.~Xu, ``Consensus-based distributed quantum kernel learning for speech recognition,'' in {\em 2025 IEEE International Conference on Acoustics, Speech, and Signal Processing Workshops (ICASSPW)}, pp.~1--5, IEEE, 2025.

\bibitem{leyton2021robust}
V.~Leyton-Ortega, A.~Perdomo-Ortiz, and O.~Perdomo, ``Robust implementation of generative modeling with parametrized quantum circuits,'' {\em Quantum Machine Intelligence}, vol.~3, no.~1, p.~17, 2021.

\bibitem{benedetti2019generative}
M.~Benedetti, D.~Garcia-Pintos, O.~Perdomo, V.~Leyton-Ortega, Y.~Nam, and A.~Perdomo-Ortiz, ``A generative modeling approach for benchmarking and training shallow quantum circuits,'' {\em npj Quantum information}, vol.~5, no.~1, p.~45, 2019.

\bibitem{chen2025differentiable}
S.~Y.-C. Chen, C.-Y. Liu, K.-C. Chen, W.-J. Huang, Y.-J. Chang, and W.-H. Huang, ``Differentiable quantum architecture search in quantum-enhanced neural network parameter generation,'' {\em arXiv preprint arXiv:2505.09653}, 2025.

\bibitem{moll2018quantum}
N.~Moll, P.~Barkoutsos, L.~S. Bishop, J.~M. Chow, A.~Cross, D.~J. Egger, S.~Filipp, A.~Fuhrer, J.~M. Gambetta, M.~Ganzhorn, {\em et~al.}, ``Quantum optimization using variational algorithms on near-term quantum devices,'' {\em Quantum Science and Technology}, vol.~3, no.~3, p.~030503, 2018.

\bibitem{chen2025learning}
K.-C. Chen, H.~Matsuyama, and W.-H. Huang, ``Learning to learn with quantum optimization via quantum neural networks,'' {\em arXiv preprint arXiv:2505.00561}, 2025.

\bibitem{liu2024quantum}
C.-Y. Liu, C.-H.~H. Yang, H.-S. Goan, and M.-H. Hsieh, ``A quantum circuit-based compression perspective for parameter-efficient learning,'' {\em arXiv preprint arXiv:2410.09846}, 2024.

\bibitem{liu2024qtrl}
C.-Y. Liu, C.-H.~A. Lin, C.-H.~H. Yang, K.-C. Chen, and M.-H. Hsieh, ``Qtrl: Toward practical quantum reinforcement learning via quantum-train,'' in {\em 2024 IEEE International Conference on Quantum Computing and Engineering (QCE)}, vol.~2, pp.~317--322, IEEE, 2024.

\bibitem{lin2024quantum}
C.-H.~A. Lin, C.-Y. Liu, and K.-C. Chen, ``Quantum-train long short-term memory: Application on flood prediction problem,'' in {\em 2024 IEEE International Conference on Quantum Computing and Engineering (QCE)}, vol.~2, pp.~268--273, IEEE, 2024.

\bibitem{liu2025quantum}
C.-Y. Liu, K.-C. Chen, Y.-C. Chen, S.~Y.-C. Chen, W.-H. Huang, W.-J. Huang, and Y.-J. Chang, ``Quantum-enhanced parameter-efficient learning for typhoon trajectory forecasting,'' {\em arXiv preprint arXiv:2505.09395}, 2025.

\bibitem{chen2025quantum}
K.-C. Chen, S.~Y.-C. Chen, C.-Y. Liu, and K.~K. Leung, ``Quantum-train-based distributed multi-agent reinforcement learning,'' in {\em 2025 IEEE Symposium for Multidisciplinary Computational Intelligence Incubators (MCII Companion)}, pp.~1--5, IEEE, 2025.

\bibitem{cao2019gcnet}
Y.~Cao, J.~Xu, S.~Lin, F.~Wei, and H.~Hu, ``Gcnet: Non-local networks meet squeeze-excitation networks and beyond,'' in {\em Proceedings of the IEEE/CVF international conference on computer vision workshops}, pp.~0--0, 2019.

\bibitem{lee2019srm}
H.~Lee, H.-E. Kim, and H.~Nam, ``Srm: A style-based recalibration module for convolutional neural networks,'' in {\em Proceedings of the IEEE/CVF International conference on computer vision}, pp.~1854--1862, 2019.

\bibitem{zhang2023qca}
J.~Zhang, P.~Cheng, Z.~Li, H.~Wu, W.~An, and J.~Zhou, ``Qca-net: Quantum-based channel attention for deep neural networks,'' in {\em 2023 International Joint Conference on Neural Networks (IJCNN)}, pp.~1--7, IEEE, 2023.

\bibitem{wierichs2022general}
D.~Wierichs, J.~Izaac, C.~Wang, and C.~Y.-Y. Lin, ``General parameter-shift rules for quantum gradients,'' {\em Quantum}, vol.~6, p.~677, 2022.

\bibitem{bergholm2018pennylane}
V.~Bergholm, J.~Izaac, M.~Schuld, C.~Gogolin, S.~Ahmed, V.~Ajith, M.~S. Alam, G.~Alonso-Linaje, B.~AkashNarayanan, A.~Asadi, {\em et~al.}, ``Pennylane: Automatic differentiation of hybrid quantum-classical computations,'' {\em arXiv preprint arXiv:1811.04968}, 2018.

\bibitem{paszke2019pytorch}
A.~Paszke, ``Pytorch: An imperative style, high-performance deep learning library,'' {\em arXiv preprint arXiv:1912.01703}, 2019.

\bibitem{mcclean2018barren}
J.~R. McClean, S.~Boixo, V.~N. Smelyanskiy, R.~Babbush, and H.~Neven, ``Barren plateaus in quantum neural network training landscapes,'' {\em Nature communications}, vol.~9, no.~1, p.~4812, 2018.

\bibitem{thanasilp2024exponential}
S.~Thanasilp, S.~Wang, M.~Cerezo, and Z.~Holmes, ``Exponential concentration in quantum kernel methods,'' {\em Nature communications}, vol.~15, no.~1, p.~5200, 2024.

\end{thebibliography}
\end{document}